# Surface tension model for surfactant solutions at the critical micelle concentration

Sergei F. Burlatsky[a], Vadim V. Atrazhev[b,c,*], Dmitry V. Dmitriev[b,c], Vadim I. Sultanov[c], Elena N. Timokhina[b,c], Elena A. Ugolkova[c], Sonia Tulyani[a], Antonio Vincitore[a]

[a] United Technologies Research Center, 411 Silver Lane, East Hartford, CT 06108, United States

[b] Russian Academy of Science, Emanuel Institute of Biochemical Physics, Kosygin str. 4, Moscow, 119334, Russia

[c] Science for Technology LLC, Leninskiy pr-t 95, 119313, Moscow, Russia

[*] Corresponding author. Fax: +7 499 137 82 31

*E-mail address:* vvatrazhev@deom.chph.ras.ru

**Abstract**

A model for the limiting surface tension of surfactant solutions (surface tension at and above the critical micelle concentration, cmc) was developed. This model takes advantage of the equilibrium between the surfactant molecules on the liquid/vacuum surface and in micelles in the bulk at the cmc. An approximate analytical equation for the surface tension at the cmc was obtained. The derived equation contains two parameters, which characterize the intermolecular interactions in the micelles, and the third parameter, which is the surface area per surfactant molecule at the interface. These parameters were calculated using a new atomistic modeling approach. The performed calculations of the limiting surface tension for four simple surfactants show good agreement with experimental data (~30% accuracy). The developed model provides the guidance for design of surfactants with low surface tension values.

# 1. Introduction

Reduction of surface/interfacial tension in aqueous solutions is one of the key functions of surfactants. In a number of practical applications such as oil extraction and fire extinction foams the minimal surface tension, $\sigma_{lim}$, is a property of interest [1]. The $\sigma_{lim}$ is the limiting surface tension that can be achieved by adding surfactant to a solvent. The smaller the value of $\sigma_{lim}$, the better is the surfactant "effectiveness" [2]. Development of modeling capability to predict $\sigma_{lim}$ as a function of the composition and structure of the surfactant molecule would enable rapid development and optimization of new surfactants. We believe that this modeling capability would clarify the thermodynamic equilibrium between surfactants in the bulk, micelle and at the surface and reveal the impact of surfactant structure and composition on the limiting surface tension.

The surface tension, $\sigma$, of the surfactant solution decreases with increase of the concentration of surfactant molecules in the system. The concentration of surfactants in the system, $C$, is defined as the total number of moles of the surfactant molecules divided by the total system volume. As $C$ increases, $\sigma$ decreases. A limitation for $\sigma$ decrease is imposed by the surfactant solubility limit. However, for the majority of surfactants the system undergoes a phase separation long before $C$ reaches the solubility limit. The surfactants form a disperse phase comprised of micelles along with the solution of individual surfactant molecules in the bulk. The concentration at which the micelles formation starts is called critical micelle concentration (cmc). For $C$ lower than the cmc, surfactants are distributed between the bulk of the solution and the surface. The bulk concentration of surfactants, $C_{bulk}$, and the surface concentration, $\Gamma$, both increase with $C$ until $C$ reaches cmc. After $C$ reaches cmc, all added 'excess' surfactants form micelles so that the bulk concentration of the individual surfactant molecules is constant and equal to the cmc. The surface concentration also reaches the saturation level, $\Gamma_{cmc}$. Since the surface tension is governed by $\Gamma$ and is not sensitive to the micelle concentration in the bulk, it also reaches the saturation level that is equal to $\sigma$ at cmc, $\sigma_{cmc}$. Thus, $\sigma_{cmc}$ is the limiting surface tension that can be archived by using the specific surfactant, i.e. $\sigma_{cmc} = \sigma_{lim}$.

A surface tension isotherm is the dependence of $\sigma$ on $\ln C$. A typical surface tension isotherm is plotted schematically in Figure 1a. At $C >$ cmc the surface tension is constant. At $C <$ cmc the isotherm may be roughly divided into two regions (see Fig. 1a). In region (I) the slope of the curve $\sigma(\ln C)$ decreases with $\ln C$ while in region (II) $\sigma$ is a linear function of $\ln C$. Surface concentration, $\Gamma$, increases in region (I). In region (II) the saturation of the surface with surfactant molecules is established and $\Gamma$ asymptotically approaches its limiting value, $\Gamma_\infty$ (see Fig 1b). The dependence of $\sigma$ on $\Gamma$ becomes very steep (Figure 1c) in region (II) in spite of the fact that the dependence of $\Gamma$ on $\ln C$ is weak in this region.

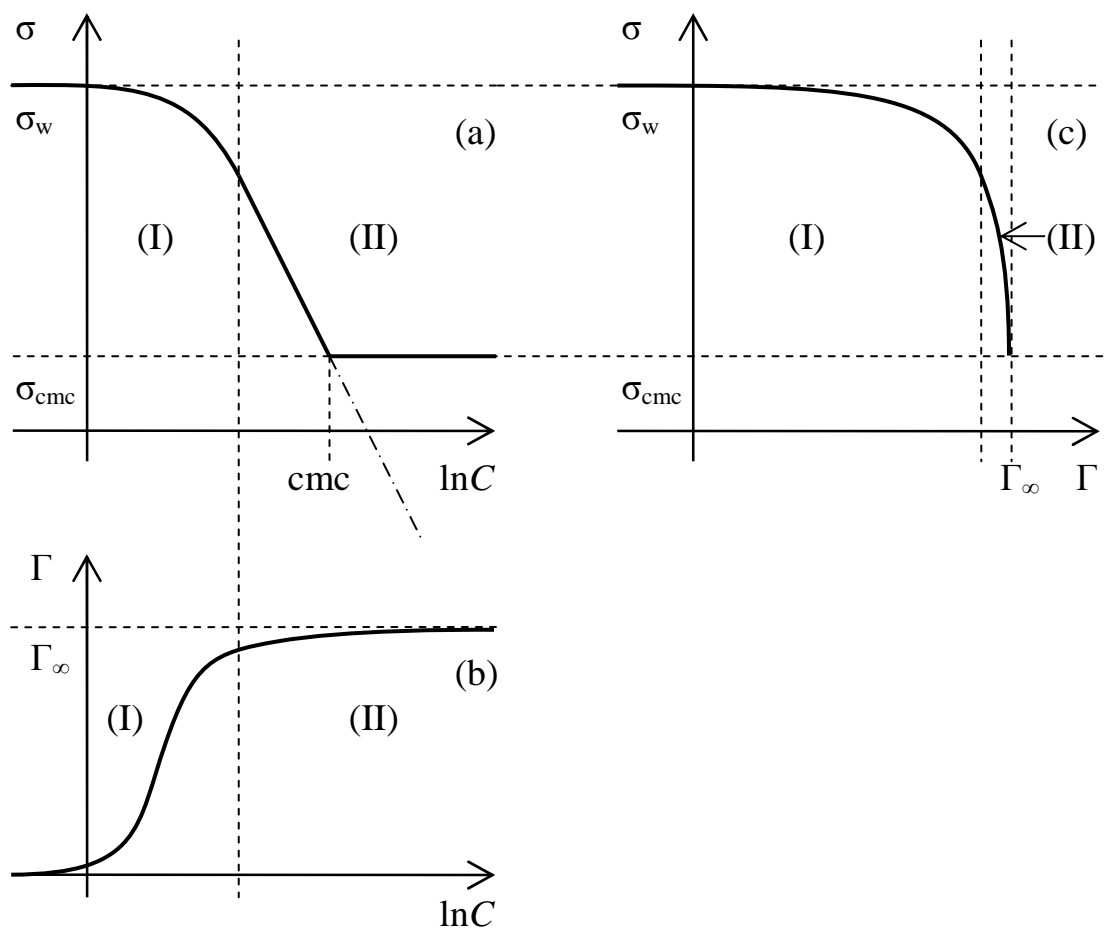

**Fig. 1.** (a) Schematic representation of typical surface tension isotherm of water/surfactant solution. (b) Surface concentration of surfactant molecules, $\Gamma$. (c) Dependence of surface tension of water/surfactant solution on $\Gamma$.

Currently several semi-empirical and molecular-level modeling approaches are available in the literature for the surface tension of aqueous solution of surfactants. The semi-empirical approach developed in [3,4] is based on Quantitative Structure-Property Relationships (QSPR) widely used in chemistry and biochemistry. QSPR is based on empirical correlations between the physical or chemical properties and the chemical structure of the compound. The chemical structure is characterized by a set of specific values called descriptors. Wang and co-workers performed QSPR analysis of the limiting surface tension of solutions of nonionic (polyethyleneoxide type) [4] and anionic [3] surfactants. According to [3], for anionic surfactants the descriptors are: the heat of formation ($\Delta H_f$, quantum chemical descriptor), dipole moment of the molecule ($D$, quantum chemical descriptor governed mainly by the polarity of the head group), and Kier & Hall index of the hydrophobic tail ($KH0$, topological descriptor governed the size of tail). According to [4], for nonionic surfactants the polarity of the head group was accounted *via* the number of oxygen atoms in the head (*NO*) as opposed to the dipole momentum, *D*, that is used as a descriptor for anionic surfactants. The models developed in [3,4] predict the dependencies of $\sigma_{cmc}$ on chemical structure of the surfactants. In addition, they predict the dependence of $\sigma_{cmc}$ on temperature and the dependence of $\sigma_{cmc}$ on concentration of added electrolyte (NaCl) for anionic surfactants. The QSPR approach demonstrates good agreement with experiment when a large experimental database for surfactants of similar structure is available. Applicability of the QSPR approach for design of new surfactant molecules is challenging when experimental database for surfactants of the similar structure to the desired molecule is not available.

Direct atomistic (Molecular Dynamics or Monte Carlo) calculation of surface effects is a challenging problem. The surface energy is small as compared to thermodynamic fluctuation of the total energy in relatively small systems available for the treatment by Molecular Dynamics or Monte Carlo. To make the results statistically significant the calculation of extremely long MD trajectories is required.

The Kirkwood-Buff (KB) molecular modeling approach [5–22] is fully atomistic. It utilizes Kirkwood-Buff relation between $\sigma$ and diagonal components of the pressure tensor [23]. Using this approach one can calculate the surface tension of water/surfactant solution as a function of surfactant surface concentration, $\Gamma$. This method predicts the correct qualitative dependence $\sigma(\Gamma)$ and reproduces other important trends. However, the direct application of this approach to calculation of $\sigma_{cmc}$ would be challenging as it would involve the calculation or measurement of the surface concentration of surfactant molecules at cmc with subsequent direct calculation of the surface tension at this point. Due to the steep dependence of $\sigma$ on $\Gamma$, a small error of about ~5% in $\Gamma_{cmc}$ would cause more than 100% error in $\sigma_{cmc}$. Indeed, calculated values of $\sigma_{cmc}$ at water/heptane interface in Kirkwood-Buff formalism in [11] correlate with variation of experimental data. However, the traditional approach overestimates the values of $\sigma_{cmc}$ by approximately 25 times (see Fig. 9 therein).

The combination of Kirkwood-Buff approach with Dissipative Particle Dynamics (DPD) was utilized in [24] to demonstrate qualitatively behavior of water/surfactants solution. DPD is a coarse-grained Molecular Dynamics approach [22,24–29]. It enables calculations on relatively long time-scales, up to microsecond, due to dramatically reduced number of particles and highly simplified inter-particles interactions. The reduction of the number of particles is achieved by grouping of several atoms into one block (united atom). The results presented in [24] demonstrate sequential transition from the bulk solution of individual surfactant molecules to micelle formation through surfactants monolayer formation at the surface when the concentration of the surfactants in the system gradually increased. The surface tension calculated by Kirkwood-Buff formula from trajectories obtained by DPD method also has a qualitatively correct dependence on the bulk concentration of the surfactant and reflects the micelles formation [24]. However, quantitative calculation of $\sigma_{cmc}$ currently cannot be obtained by DPD because the interactions used in the DPD are oversimplified.

In this paper, we develop a new method for prediction of $\sigma_{cmc}$ for surfactant solution. The proposed approach takes into account the thermodynamic equilibrium between the surfactant molecules at the surface and in the micelles that establishes at and above the cmc point. We expressed $\sigma_{cmc}$ through the limiting value of the surface concentration at the surface, $\Gamma_{cmc}$, and the change of the free energy, $\Delta F$, in the transfer process of one surfactant molecule from the micelle to the surface (see Fig. 2). The $\Delta F$ is equal to the difference of the free energy per one surfactant molecule in the micelle and at the surface. The atomistic calculation of $\Delta F$ would require prohibitively long computational time. An approximate equation for $\sigma_{cmc}$ was obtained using additional approximations. Parameters of this equation were calculated for four simple structure surfactants using a new atomistic approach. Predicted values of $\sigma_{cmc}$ are in a reasonable agreement with experimental data.

The rest of the paper is organized as follows. The derivation of exact analytical expression for the surface tension at cmc is presented in Section II. The derivation of approximate expression the surface tension at cmc is presented in Section III. Atomistic approach for the calculation of the model parameters is presented in Section IV. The modeling results for four selected surfactants are presented in Section V. The model predictions are discussed in Section VI.

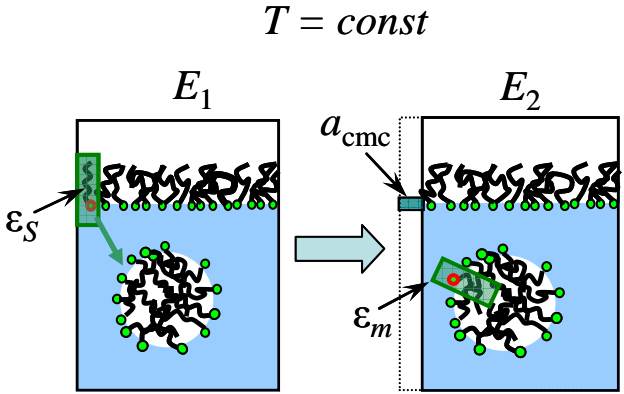

**Fig. 2.** Schematic representation of reduction of the surface area and transfer of small Subsystem containing one surfactant molecule and the adjacent water molecules from the surface to the micelle Subsystem is highlighted in green.

**2. Surface tension at cmc**

In this Section, we derive the thermodynamic expression for $\sigma_{cmc}$. This expression presents $\sigma_{cmc}$ as a function of free energy difference between the surfactant molecule in the micelle and that at the surface and of the area per one surfactant molecule at the surface. The above free energy difference is equal to the free energy gained in course of the quasi-equilibrium decrease of the surface area with simultaneous transfer of a surfactant molecule from the surface into the micelle, as indicated in Fig. 2.

The surface tension of the surfactant solution is equal to the change of the free energy that results from the isothermal surface area reduction normalized by the exercised change of the surface area. This reduction is accomplished by the transfer of a small Subsystem highlighted in green in Fig. 2 from the surface to the micelle. This Subsystem contains a surfactant molecule with the adjacent water molecules interacting with surfactant molecule. The base surface of the transferred Subsystem is equal to the change of the surface area (see Fig. 2). To calculate the limiting surface tension we calculate the change in free energy caused by the isothermal reduction of the surface area for the water/surfactant solution with the total surfactant concentration equal to or higher than cmc. At such concentrations, the surfactant molecules at the surface, in the bulk of the solution and in the micelles are in thermodynamic equilibrium. In the actual solution, the surfactant molecules would be transferred from the surface to the micelles through the bulk of the solution. However, we calculate thermodynamic potential change for the direct transfer of the surfactant molecule from the surface to the micelle taking advantage of the triple thermodynamic equilibrium.

In the system under consideration, the surface concentration equals to the limiting surface concentration, $\Gamma_{cmc} = 1/a_{cmc}$. Here $a_{cmc}$ is the limiting value of the area of the liquid/gas surface per one surfactant molecule. We consider the isothermal quasi-equilibrium reduction of the surface area by $a_{cmc}$ that is much smaller than the total surface area of the system. That reduction is achieved by the transfer of Subsystem from the surface to the micelle, as indicated in the Fig. 2. The total micelles/water interface area increases by the surface area per one surfactant in the micelle while the micelle/water interface concentration of surfactants remains constant. The surface concentration of surfactants also does not change. The change of the energy of the system in course of isothermal transfer process is

$$\Delta E = \varepsilon_m - \varepsilon_S \tag{1}$$

Here $\varepsilon_S$ and $\varepsilon_m$ are the energies of the Subsystem on the surface and in the micelle, correspondingly. In the transfer process, we changed the entropy of the system. Therefore, to keep the temperature constant we should add heat to the system. This heat is equal to

$$\Delta Q = T(s_m - s_s) \tag{2}$$

Here $s_S$ and $s_m$ are the entropy of Subsystem at the surface and in the micelle, correspondingly.

According to the first principle of thermodynamics, the change of the energy of the system is equal to the heat added to the system minus the work done by the system:

$$\Delta E = \Delta Q - W \tag{3}$$

The work done by the system is equal to the change of the surface energy of the system:

$$W = \sigma_{cmc} a_{cmc} \tag{4}$$

Substituting Eqs. (1), (2) and (4) into Eq. (3) we obtain

$$\varepsilon_m - \varepsilon_S = T(s_m - s_S) - \sigma_{cmc} a_{cmc} \tag{5}$$

Solving Eq. (5) with respect to $\sigma_{cmc}$ we obtain

$$\sigma_{cmc} = \frac{(\varepsilon_S - \varepsilon_m) - T(s_S - s_m)}{a_{cmc}} = \frac{f_S - f_m}{a_{cmc}} \tag{6}$$

Here f is a free energy per one surfactant, $f = \varepsilon - Ts$. Eq. (6) is a microscopic implementation of general thermodynamic equation for the surface tension:

$$\sigma = \left(\frac{\partial F}{\partial a}\right)_{NVT} \tag{7}$$

The derived analytical expression (6) for $\sigma_{cmc}$ contains two contributions having clear physical meanings. The first, energetic, term is equal to the difference between the surfactant energy at the surface and in the micelle. This term is governed mostly the density of interaction between the opposite layers in the micelle, which is indicated by green slab in Fig. 3b. The second, entropic, term is equal to the difference between the entropy of the surfactant at the surface and in the micelle, multiplied by the temperature. This term is governed mostly by the restrictions in the mobility of the surfactant molecules in the micelle imposed by additional intermolecular interaction in the micelle.

The atomistic calculation of the entropic contribution in the free energy difference requires prohibitively long computational time [30]. In the next section, the entopic term in Eq. (6) is approximate expressed through the effective mechanical property of the micelle that can be more easily calculated through atomistic modeling.

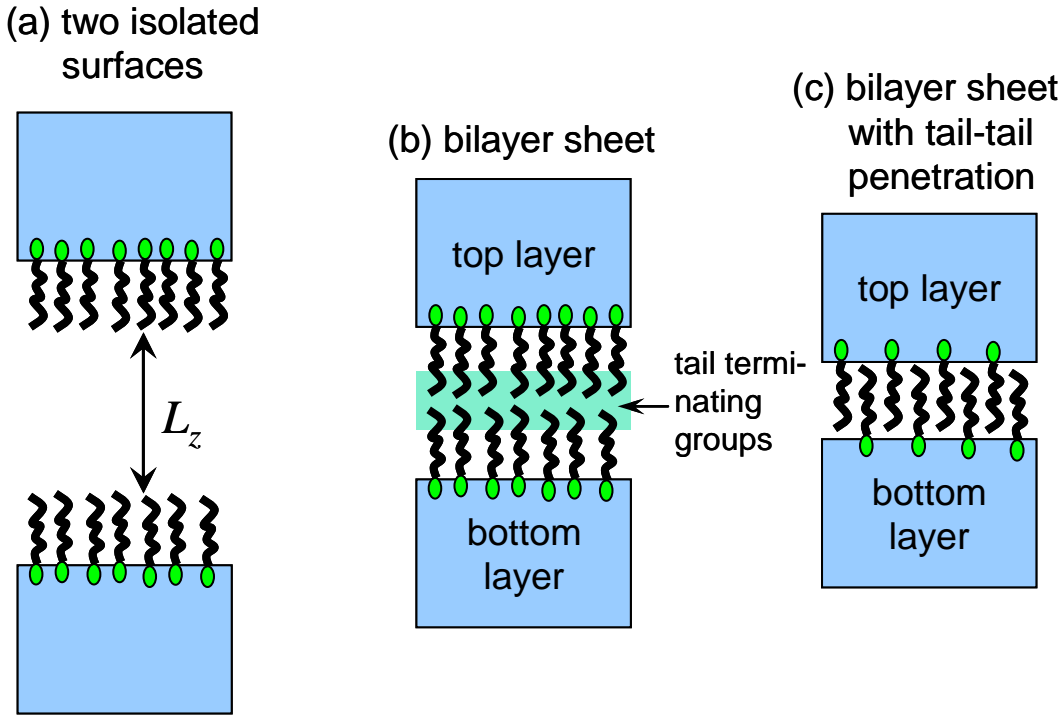

**Fig. 3.** Schematic representation of: (a) two isolated surfaces (liquid/vacuum interfaces), and bilayer (b) without and (c) with cross-micelle tail-tail penetration.

**3. Approximate equation for the surface tension at cmc**

In this section, we derive the approximate equation for the surface tension at the cmc. This equation neglects the effects of the curvature of the micelle using the bilayer sheet (see Fig. 3b) approximation. Taking advantage of the similarity of the surfactant configurations on the surface and in the bilayer, we derived analytical equation for $\sigma_{cmc}$. This equation expresses $\sigma_{cmc}$ through $a_{cmc}$ and $\Delta\varepsilon$, similarly to Eq. (6). However, the entropic term is expressed though parameter of interaction of the surfactants in the micelle.

To obtain the closed analytical equation for $\sigma_{cmc}$, we neglect the micelle shape effects utilizing the bilayer sheet approximation. The bilayer sheet is two infinitely long parallel water surfaces filled by surfactant molecules. The surfactant tails are in the contact in the bilayer, as shown in Fig. 3b. The bilayer is a limiting case of a vesicle of a large diameter. The optimal surfactant configuration in the micelle is more complicated. The different types of surfactants form the micelles of different size and shape (spheres, ellipsoids, cylinders etc.) [2]. Moreover, the variety of subsequent changes in micelle geometry was observed while changing the surfactant concentration above the cmc [2]. However, the observed variations of surface tension above the cmc were negligible. This implies that the free energy of the surfactant molecule in the micelle does not change substantially as a function of the micelle geometry and to some extend justifies our approximation. The free energy of bilayer is higher or equal to the actual micelle free energy. Therefore, our analytical approximation provides a lower bound for $\sigma_{cmc}$.

In the bilayer sheet approximation Eq. (6) reads:

$$\sigma_{cmc} = \frac{F_M - F_S}{2Na_{cmc}} \tag{8}$$

Here N is the number of surfactant molecules in each layer of the bilayer, $F_S$ is the free energy of two isolated surfaces (bilayer in the limit of infinite thickness) and $F_M$ is the free energy of the bilayer at optimized thickness.

The free energy $F$ is expressed through the partition function $Z$ by the well-known equation:

$$F = -k_B T \ln Z \tag{9}$$

Thus, the surface tension at the cmc is:

$$\sigma_{cmc} = \frac{k_B T}{2Na_{cmc}} \ln\left(\frac{Z_S^2}{Z_M}\right) \tag{10}$$

where $Z_S$ is the partition functions of the isolated surface containing $N$ surfactants and $Z_M$ is the partition functions of the bilayer containing $2N$ surfactants.

The partition function of the system of surfactants at the surface (in the micelle) is by definition:

$$Z = \sum_n e^{-E(n)/k_B T} \tag{11}$$

Here $n$ denotes the configuration of the surfactants at the surface (in the micelle) and the sum is taken over all possible configurations.

To simplify calculation of Z we assume that the surfactants form a periodic lattice with equal lattice spacing both at the surface and in the bilayer. We also assume that the major difference in the entropy of the surfactants at the surface and in the bilayer comes from the restriction of the surfactant motion in through-plane direction in the bilayer. Coordinate $\varphi_n$ of the surfactant number n is defined in Fig. 4. This coordinate is a deviation of the surfactant head from the water surface. In our model, the configurations of the surfactants on the surface are defined by their position $\varphi_n$ relative to the surface.

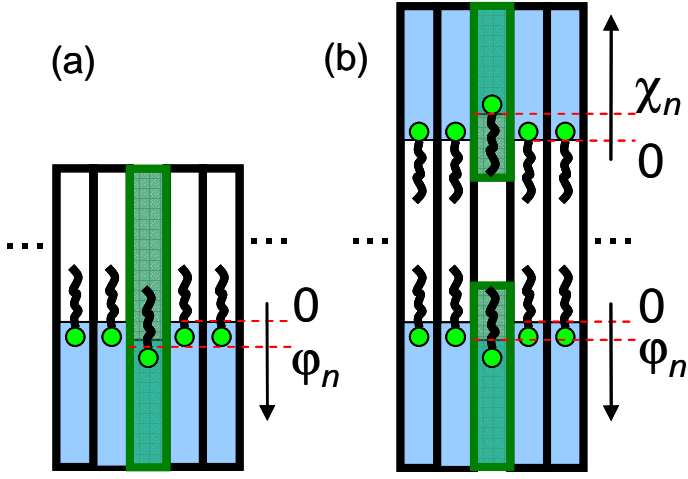

**Fig. 4.** Configurations of surfactants (a) on liquid/vacuum surface and (b) in the bilayer.

The partition function of the surfactants on the surface is:

$$Z_S = \int_{-\infty}^{\infty} \prod_n d\varphi_n \exp\left(-\frac{E_S\{\varphi\}}{k_B T}\right) \tag{12}$$

The integration in Eq. (12) is performed over positions $\varphi_n$ of the surfactants at the surface. In the optimal (ground state) configuration, all surfactants are located in one plane, as shown in Fig. 4. In this configuration $\varphi_n = 0$ for $n = 1...N$. The surface ground state energy is $E_{S0}$. In the nearest neighbors interaction approximation with parabolic potentials the energy of the surface configuration is

$$E_S\{\varphi\} = E_{S0} + \frac{K}{2} \sum_{\langle nm \rangle} (\varphi_n - \varphi_m)^2 \tag{13}$$

Here K is a 'stiffness' parameter and $\langle nm \rangle$ means summation over nearest neighbor surfactants.

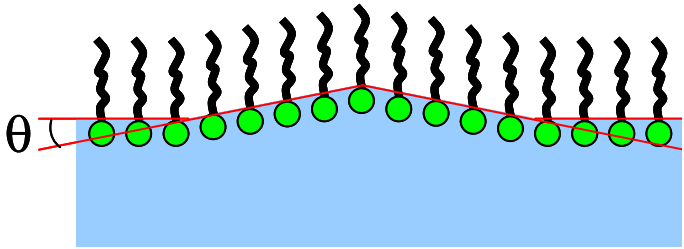

**Fig. 5.** Schematic representation of weak perturbation in the form of triangle wave with small angle θ on the surface and the corresponding configuration of surfactant molecules.

The stiffness parameter $K$ is calculated as follows. We consider a small perturbation of the surface in the form of triangular wave with small angle $\theta$, as shown in Fig. 5. For two nearest surfactants $n$ and $m$ located at the perturbation wave, the difference of the coordinates $\varphi$ is $\varphi_n - \varphi_m = l \sin\theta$, where $l = \sqrt{a_{cmc}}$ is the mean distance between the surfactants at the surface. Substituting $\varphi_n - \varphi_m$ into Eq. (13), we obtain the deviation of the energy from the ground state normalized by a number of the surfactants in the perturbation wave

$$\delta\varepsilon = \frac{K a_{cmc} \sin\theta^2}{2} \approx \frac{K a_{cmc} \theta^2}{2} \tag{14}$$

This triangular perturbation wave causes the increase of the surface area, $\delta A$, which results in the additional surface energy, $\delta\varepsilon = \sigma_{cmc} \delta A$. The increase of the surface area per one surfactant in the perturbation wave is $\delta A = a_{cmc} \frac{1-\cos\theta}{\cos\theta}$, and additional surface energy per one surfactant in the perturbation wave is equal to

$$\delta\varepsilon = \sigma_{cmc} a_{cmc} \frac{1-\cos\theta}{\cos\theta} \approx \frac{\sigma_{cmc} a_{cmc} \theta^2}{2} \tag{15}$$

Equating the right hand sides of Eqs. (14) and (15) we conclude that, the surface stiffness, $K$, is equal to the surface tension $\sigma_{cmc}$:

$$K = \sigma_{cmc} \tag{16}$$

Energy functional (13) is diagonalized by Fourier transformation

$$\varphi_n = \sum_k \varphi_n e^{i k \cdot n} \tag{17}$$

which results in

$$E_S\{\varphi\} = E_{S0} + \sigma_{cmc} \sum_k \varepsilon(k) \varphi_k \varphi_{-k} \tag{18}$$

Here the sum is over all wave vectors $\mathbf{k} = (k_x, k_y)$ in the Brillouin zone and the spectrum $\varepsilon(k_x, k_y)$ depends on the lattice type formed by the surfactants. For example, the spectrum for the triangular lattice is[1]:

---

[1] The results for the $\sigma_{cmc}$ weakly depend on the choice of the lattice (the difference is less than 2mN/m). Therefore, we use the spectrum for the triangular lattice.

$$\varepsilon(k_x, k_y) = 3 - \cos k_x - 2\cos\frac{k_x}{2}\cos\frac{\sqrt{3}k_y}{2} \qquad (19)$$

Substituting Eq. (13) into (12), we calculate the partition function explicitly:

$$Z_S = e^{-E_{S0}/kT} \prod_{k_x,k_y} \sqrt{\frac{k_B T}{\sigma_{cmc}\varepsilon(k_x,k_y)}} \qquad (20)$$

The partition function of the bilayer comprising of 2N surfactants is

$$Z_M = \int_{-\infty}^{\infty} \prod_n d\varphi_n d\chi_n \exp\left(-\frac{E_M\{\varphi,\chi\}}{kT}\right) \qquad (21)$$

Here $\varphi_n$ is the deviation ~~of the center mass~~ of the surfactant head from the ground-state position of the $n$-th surfactant of the bottom surface of the bilayer and $\chi_n$ is the deviation of the center mass of the surfactant head from the ground-state position for the top surface of the bilayer (see Fig. 5). In the ground state of the bilayer, $\varphi_n = \chi_n = 0$. The energy of the surfactants in the bilayer differs from that at the surface by the value of the cross-micelle interaction $U_{cr}$ which is a function of the distance between the opposite surfactants of top and bottom surfaces ($\varphi_n - \chi_n$):

$$E_M\{\varphi,\chi\} = 2E_{s0} + \frac{\sigma_{cmc}}{2}\sum_{\langle nm\rangle}\left[(\varphi_n - \varphi_m)^2 + (\chi_n - \chi_m)^2\right] + \sum_n U_{cr}(\varphi_n - \chi_n) \qquad (22)$$

The dependence $U_{cr}(r)$ is calculated numerically using Molecular Mechanics approach, described in section IV. The typical dependence $U_{cr}(r)$ is shown in Fig. 6. Near the optimal configuration $U_{cr}(r)$ is parabolic:

$$U_{cr}(\varphi - \chi) = -\Delta\varepsilon + \frac{\kappa(\varphi - \chi)^2}{2} \qquad (23)$$

Here $\Delta\varepsilon = \varepsilon_s - \varepsilon_m$ is the difference between the energy of one surfactant molecule in the ground state at the surface and in the bilayer, normalize by $N$, and $\kappa$ is the bilayer stiffness. Performing Fourier transformation for Eq. (22) and substituting the diagonalized energy function into Eq. (21) we calculate the partition function of the bilayer explicitly:

$$Z_M = e^{-(2E_S - N\Delta\varepsilon)/k_B T} \prod_{k_x,k_y} \sqrt{\frac{k_B T}{\sigma_{cmc}\varepsilon(k_x,k_y)}} \sqrt{\frac{k_B T}{\kappa + \sigma_{cmc}\varepsilon(k_x,k_y)}} \qquad (24)$$

Substituting the partition functions (20) and (24) into Eq. (10) we obtain the expression for $\sigma_{cmc}$:

$$\sigma_{cmc} = \frac{\Delta\varepsilon}{2a_{cmc}} - \frac{1}{2a_{cmc}} k_B T \iint_{BZ} \frac{dk_x dk_y}{V_{BZ}} \ln\left(1 + \frac{\kappa}{\sigma_{cmc}\varepsilon(k_x,k_y)}\right) \quad (25)$$

Here we approximate the summation over $k_x$ and $k_y$ by the integration over Brillouin zone ($V_{BZ}$ is Brillouin zone volume). According to Eq. (25) the surface tension is governed by the surface area per one surfactant molecule $a_{cmc}$ and two parameters of the cross-micelle interaction, $\Delta\varepsilon$ and the bilayer stiffness $\kappa$.

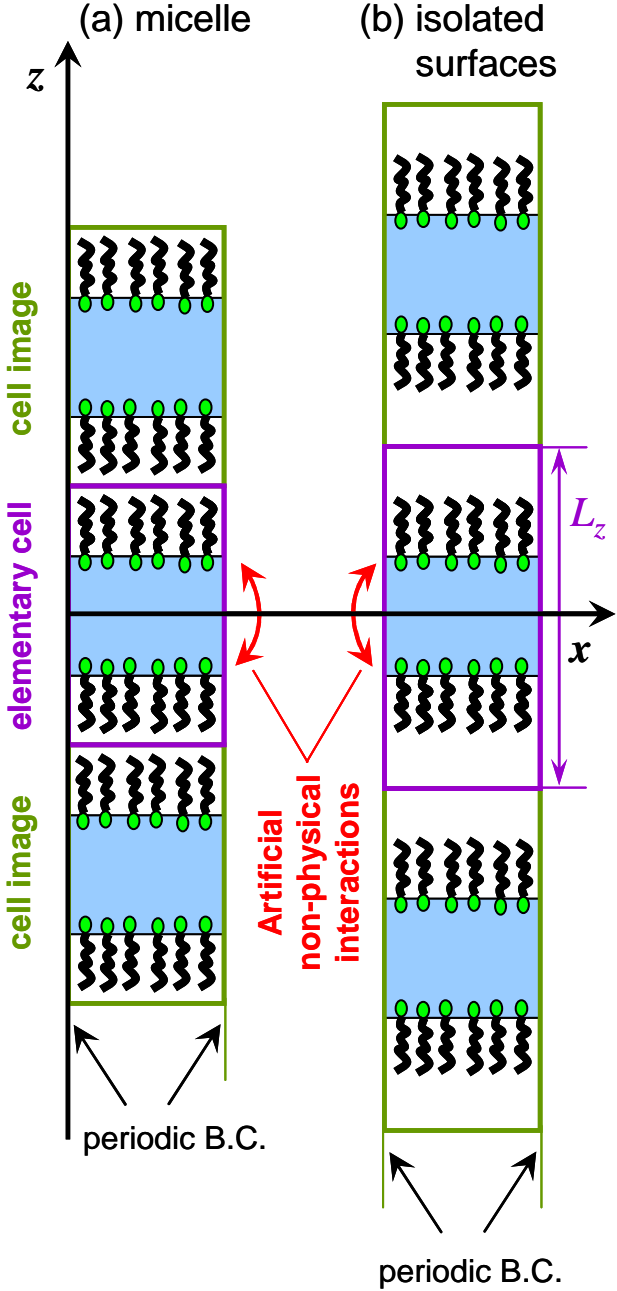

**Fig. 6.** Numerical implementation of the model. We used periodic boundary conditions in *x*, *y* and *z* directions. The micelle model (a) was generated from the surface model (b) by decrease of $L_z$. Artificial non-physical interactions are indicated by the red arrowhead line.

The leading contribution to the integral in Eq. (25) is governed by the modes with small wave vectors $\mathbf{k}$. Expanding $\varepsilon(k_x,k_y)$ near $\mathbf{k}=0$ and performing integration in Eq. (25) we obtain the following approximate equation for $\sigma_{cmc}$

$$\sigma_{cmc} = \frac{\Delta\varepsilon}{2a_{cmc}} - \frac{k_B T}{2a_{cmc}}\frac{\pi}{2}\left[\frac{\kappa}{\pi^2\sigma_{cmc}}\ln\left(1+\frac{\pi^2\sigma_{cmc}}{\kappa}\right)+\ln\left(1+\frac{\kappa}{\pi^2\sigma_{cmc}}\right)\right] \qquad (26)$$

The first terms in the right hand sides of Eq. (25) and Eq. (26) are the energy density of the cross-micelle interaction. The second, entropic, terms are proportional to the temperature. The entropic contribution to σcmc is always negative because the cross-micelle interaction imposes additional restriction on the number of configurations of the surfactants in the micelle. The magnitude of this restriction is governed by the bilayer stiffness parameter $\kappa$.

Eq. (26) for σcmc illuminates the background physics and specifies the critical parameters that govern the surface tension. The main contribution to the surface tension comes from the energy density of the cross-micelle interaction. In the micelles with low penetration of the tail, see Fig. 3b, the cross-micellar interaction is governed by the interaction between the tail terminating groups of the top and the bottom layers of the bilayer. The cross-micelle interaction for the surfactants with long enough tail is mostly determined by the van der Waals cross-micelle 'tail-tail' interaction between the tail terminating groups belonging to the top an the bottom layers of the bilayer. Therefore, the surfactants with weakly interacting termination groups of tails should demonstrate low surface tension at cmc. Hydrocarbon and fluorocarbon tails with $CH_3$ and $CF_3$ terminating groups are typical for surfactant molecules. Both $CH_3$ and $CF_3$ terminating groups have low interaction energy. However, the fluorocarbon tails with $CF_3$ terminating groups have lower interaction energy per unit area than the hydrocarbon tails with $CH_3$ terminating groups. According to Eq. (26) the fluorocarbon surfactant would have smaller surface tension than the hydrocarbon surfactant of the same structure, which agrees with experimentally observed trends.[2]

In the micelles with high penetration of the tails, see Fig. 3c, $\sigma_{cmc}$ is larger than $\sigma_{cmc}$ in the system without the tails penetration (Fig. 3b). In this case, the tails penetration increases the cross-micelle interaction. We speculate that for the surfactants with the tails effective diameter (thickness) larger than the effective diameter of the head the tails penetration in the micelles is unlikely. Note that the effective diameter of the ionic head includes the size of solvated shell of water molecules.

## 4. Atomistic approach for calculation of $\sigma_{cmc}$

In this section we propose a new atomistic approach, Iterative Cycling-Optimization (ICO), and apply it to calculation of the parameters for Eq. (26) for $\sigma_{cmc}$. The ICO involves combination of Molecular Dynamics (MD) and Molecular Mechanics (MM) methods. It was implemented in the following calculation algorithm. As the first step, we prepared the model of two isolated surfaces, which consists of water slab and surfactant molecules at each water/vacuum interface. The surface model contains two layers of vacuum (~200 Å) above and below the water slab. The calculations were performed in periodic boundary conditions in x, y and z directions. In periodic system, the vacuum layers separate the water slab from its top and bottom images, see Fig. 6. The surface area per one surfactant molecule, $a_{cmc}$, was calculated by MD in periodic boundary conditions in $NP_{xy}L_zT$ ensemble (barostatting applied only to $x$ and $y$ dimensions). After that, the typical MD configuration was fixed and the bilayer model was prepared by incremental reduction of the vacuum layer thickness, $L_z$, with MM optimization of the system geometry at each step. To obtain the smooth dependence of the energy of the system, $E(L_z)$, the cycling of $L_z$ was performed. The parameters $\Delta\varepsilon$ and $\kappa$ of Eq. (26) were calculated from the obtained smooth dependence $E(L_z)$. Below we present more detail description of the ICO and justify the ICO application to $\sigma_{cmc}$ calculation.

In our approach, geometrical similarity of the bilayer and the surface was maintained in the simulations. That substantially simplifies the atomistic calculation of $\Delta\varepsilon$ and $\kappa$. Two isolated surfaces are obtained from the bilayer by extension the distance $d$ between the layers of the bilayer, see Fig. 6. When the distance $d$ becomes large enough, the interaction between the layers vanishes and the bilayer converts into two isolated surfaces. In conventional approach, one would run MD simulation of the bilayer and isolated surfaces models. The average energies of the isolated surfaces and bilayer would be calculated as the energies of the models averaged over trajectory. Subtracting these energies, one would obtain $\Delta\varepsilon$. However, because of fundamental reasons that are discussed below this procedure would require prohibitively long computational time/extensive CPU resources to obtain reliable value of $\Delta\varepsilon$. The required computational time at 8 CPU workstation is about 6 months for simple linear surfactant molecule. Arguably, the time/resources required for that type of calculations are comparable with that required for the synthesis of a new surfactant molecule. That justifies our attempt to develop new approach for the fast analytical screening of the possible candidates for the synthesis of new surfactant molecules with low $\sigma_{cmc}$.

The challenges of conventional MD approach are related to the fact that the contribution of the surface to the system properties such as pressure and energy is small as compared to bulk. Therefore, the bulk fluctuations can mask the surface tension effects in atomistic models with realistic number of particles that can be archived with reasonable CPU resources/computational time. More specifically, $\Delta\varepsilon$ is smaller than the thermodynamic fluctuation of the total energy of the system. To illustrate that, we plotted the cartoon of the hypothetic potential energy of the bilayer as a function of a phase space coordinate, see solid line in Fig. 7. The space phase coordinate symbolizes the coordinates of the surfactant and water atoms of each layer that composes bilayer relative to the centers of mass of these layers. The distance between the centers of mass of the layers optimizes the energy of the bilayer at given phase space coordinate. The dashed line in Fig. 7 shows the potential energy of two isolated surfaces that were obtained from the bilayer by extension the distance between the layers centers of mass at fixed phase space coordinate. The energy of two isolated surfaces differs from the energy of the bilayer at fixed phase space coordinate by the energy of the cross-micelle interaction, $\Delta\varepsilon$. The magnitude of $\Delta\varepsilon$ is approximately equal to 3 kcal/mol for a typical surfactant, see Table 1. The major contribution to the total energy of the system comes from the intermolecular interaction of the water molecules (water subsystem). The thermodynamic fluctuation of the water subsystem, i.e. variation of the water molecules coordinates, results in substantial time variations of the total energy. The system energy landscape is illustrated in Fig. 7. The energy difference, $\Delta E$, (15 kcal/mol per one surfactant molecule in our calculations) between two energy minima indicated in Fig. 7 is much larger than $\Delta\varepsilon$. The water subsystem resides relatively long time, $t_r \sim 10$ ps, in one minimum **A**, and then jumps into another energy minimum **B**, see Fig. 7. In real world, both surface and micelle explore all thermodynamically available local minima and the average $\Delta\varepsilon$ is equal to the difference between the average energy of the micelle and average energy of the surface. Ideally, MD simulations should explore the majority of thermodynamically available local minima. However, that would require prohibitively long computational time/extensive CPU resources.

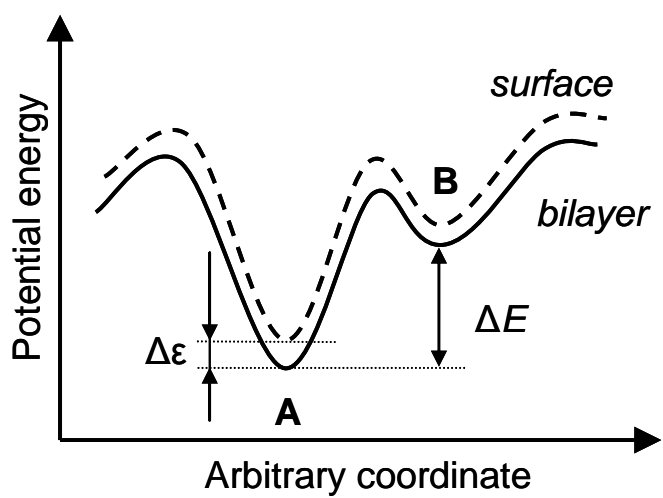

**Fig. 7**. Cartoon of the potential energy landscape of two isolated surfaces (dashed line) and bilayer (solid line). Two minima (A and B) differentiate from each other by configurations of the water molecules. At the fixed arbitrary coordinate the isolated surfaces and bilayer have the same configurations of the water molecules and differentiate from each other only by the distance between the layers, $L_z$. The energy difference between the isolated surfaces and bilayer, $\Delta\varepsilon$, which governs the $\sigma_{cmc}$, is much smaller than the energy difference between minima A and B, $\Delta E$.

To estimate required computational time we consider Gaussian distribution of the local minima over energy with standard deviation $\Delta E$. In this case, the standard deviation, $\delta E$, of the calculated average energy, $E$, is related to the number of the local minima explored by the MD trajectory during simulation by the equation

$$\frac{\delta E}{\Delta E} \approx \frac{1}{\sqrt{N}} \qquad (27)$$

To calculate $\Delta \varepsilon$ with accuracy of 10% by conventional MD approach, one should calculate the average energies of the bilayer and isolated surfaces with standard deviation $\delta E = 0.1 \cdot \Delta \varepsilon / 2 \approx 0.15$ kcal/mol. Using Eq. (27) and the value of $\Delta E \sim 15$ kcal/mol we obtain that $N \approx 10^4$. The time length of such trajectory is

$$t \approx t_r \cdot N \approx 100 \text{ ns} \qquad (28)$$

The MD modeling of such trajectory of the system contained 8 surfactant molecules and 200 water molecules at 8 CPU workstation required approximately 6 months.

We assume that the impact of the interactions between the water molecules in the bulk of the bilayer on $\Delta \varepsilon$ is relatively small and that $\sigma$ cmc is governed mostly by the energy of cross-micelle interaction. Therefore, we calculate the difference between the energy of the bilayer and isolated surfaces using ICO procedure, which keeps the bilayer and isolated surfaces in the same local minimum. Unfortunately, implementation of such procedure in MD framework is challenging, because the bilayer and isolated surfaces are treated independently in MD simulations. That motivates the choice of MM. The ICO procedure consists in multiple stepwise reduction/expansion of the vacuum layer thickness, $L_z$, with MM optimization of the system geometry at each step. The configuration of the atoms within water-surfactants layer does not change in course of the change of the system size in $z$ direction. Only the distance between the layer and its nearest replica is being changed. The MM optimization of the system geometry is minimization of the potential energy of the system through variation of positions of atoms at constant system dimensions. The multiple repetition of the reduction/expansion of Lz is required to drive the system into the deep local minimum, in which the system resides independently on Lz. In this minimum, the change of Lz effects only cross-micelle interaction. The details of calculation algorithm are presented below.

MM and MD simulations were performed using the LAMMPS software [31] with Dreiding force-field [32]. We modeled the systems containing 8 surfactant molecules, 8 counter-ions (Na$^+$) and 200 water molecules. The water layer thickness varied from 40 to 70 Å depending on the surface area per one surfactant molecule. We modeled the layer of water molecules (parallel to $xy$ plane) with periodic boundary conditions in $x$ and $y$ directions (see Fig. 6). Vacuum layers were located below and above the water layer in $z$ direction. The surfactant monolayers were located at both water-vacuum interfaces. Periodic boundary conditions in $z$ direction were used with the box size in z direction equal to $L_z$. Thus, we use periodic boundary conditions in all directions. At sufficiently large $L_z$, the model system represents two isolated surfaces as shown in Fig. 6b. At small $L_z$ it represents the bilayer as shown in Fig. 6a.

The surface area per one surfactant molecule, $a_{cmc}$, was calculated using Molecular Dynamics simulation of the isolated surfaces model (Fig. 6b) in periodic boundary conditions in $NP_{xy}L_zT$ ensemble (barostatting applied only to $x$ and $y$ dimensions). The value of $a_{cmc}$ was calculated as a product of the system sizes in x and y directions, averaged over the system trajectory, divided by the number of the surfactant molecules at one water/vacuum interface. The typical configuration of the surfactants and water molecules was selected from the trajectory obtained in MD calculation. The typical configuration was defined as the configuration with the surface area that is most close to the average surface area. The dimensions of the system in x and y directions were fixed and we obtained the MM model of two isolated surfaces.

The bilayer was generated from the isolated surfaces by MM optimization of $L_z$ using ICO procedure. The distance between the uppermost atom in the layer and the lowermost atom in the replica, $d$, is depicted in the Fig. 6. At the beginning of ICO when the system models the surface, $d$ is high, approximately equal to 200 Å. After initial geometrical optimization performed at this configuration, $L_z$ was decreased so that $d$ changed from approximately 200 Å to 2 Å. Subsequently we performed $N$ steps of ICOP process decreasing $L_z$ by 0.2 Å at each step, $N = 0.5h/(0.2$ Å$)$, where $h$ is the tail length. After $N$ steps, $L_z$ decreases by $h/2$. Then the reverse sweep was performed. The system expanded in $z$ direction in $N$ steps by 0.2 Å increments at each step until $L_z$ reached initial value of approximately 200 Å. Geometrical optimization was performed after each change of the system size in $z$ direction using conjugate gradient algorithm in LAMMPS. Direct and reverse sweeps comprise a full ICO cycle.

As a result of the ICO process we obtained the energy of the system, $E$, as a function of the distance $L_z$. The typical dependencies $E(L_z)$ is illustrated in Fig. 8a by $E(L_z)$ curve for sodium perfluorooctanoate (PFO). The conversion of the energy calculations are illustrated in Fig. 8b, where the energies of the bilayer, isolated surfaces and $\Delta\varepsilon$ are plotted vs. the cycle number for PFO. At the first cycles the potential energy curves $E(L_z)$ in the forward and reverse sweeps are substantially different. This difference gradually decreases with increase the cycle number. That indicates that the system approaches the ground of the local minimum, and the change of $L_z$ results only in the change of the cross-micelle interaction. The cycling was performed until the steady state behavior of $E(L_z)$ was reached. In steady state, the cycle-to-cycle deviation of the energy was smaller than 0.2 kcal/mol for each $L_z$ during 20 consequent cycles. From the calculated $E(L_z)$ we extracted the parameter $\Delta\varepsilon$, as a difference between the energy for the largest value of $L_z$ and the energy in the minimum of $E(L_z)$. The stiffness parameter $\kappa$ is defined in Eq. (23) as a coefficient in the leading quadratic term of the expansion of cross-micelle energy, $E(L_z)$, in a Taylor series near the energy minimum. We obtained the stiffness parameter $\kappa$ by fitting the parabolic potential to the calculated dependence $E(L_z)$.

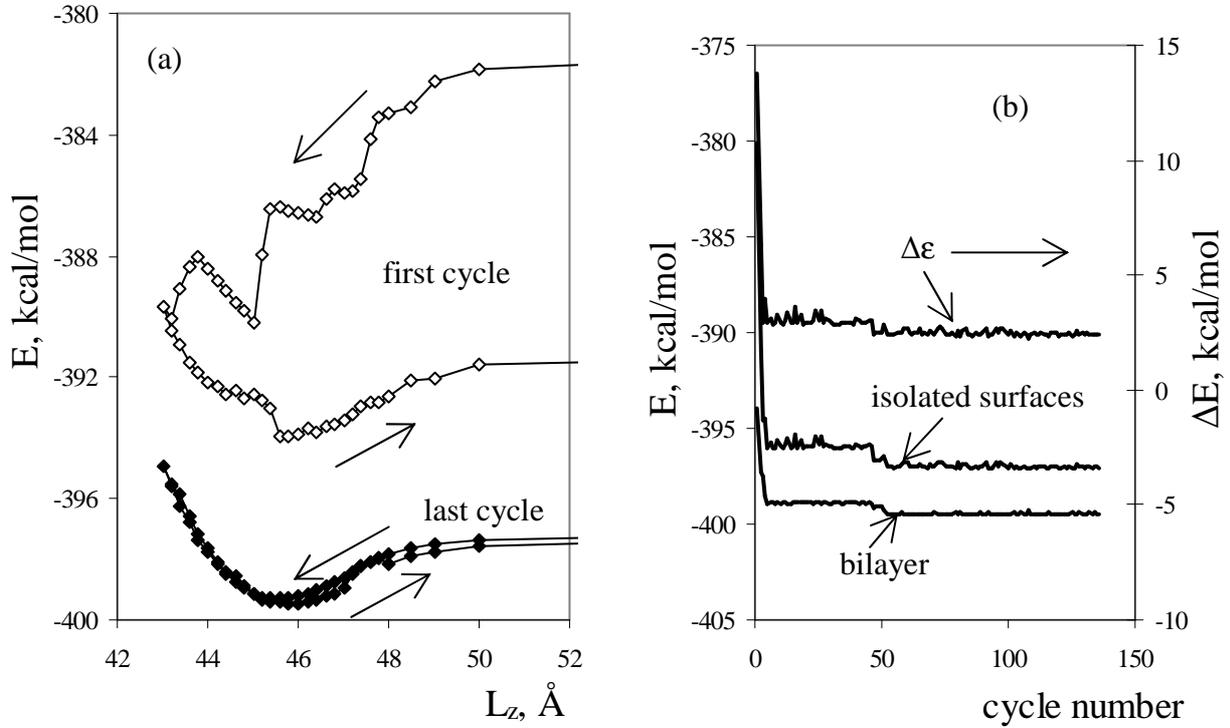

**Fig. 8.** (a) The energy of the system as a function of $L_z$ at the first and the last cycles and (b) the calculated energy of the "surface", "micelle" and the energy difference $\Delta\varepsilon$ for sodium perfluorooctanoate (PFO).

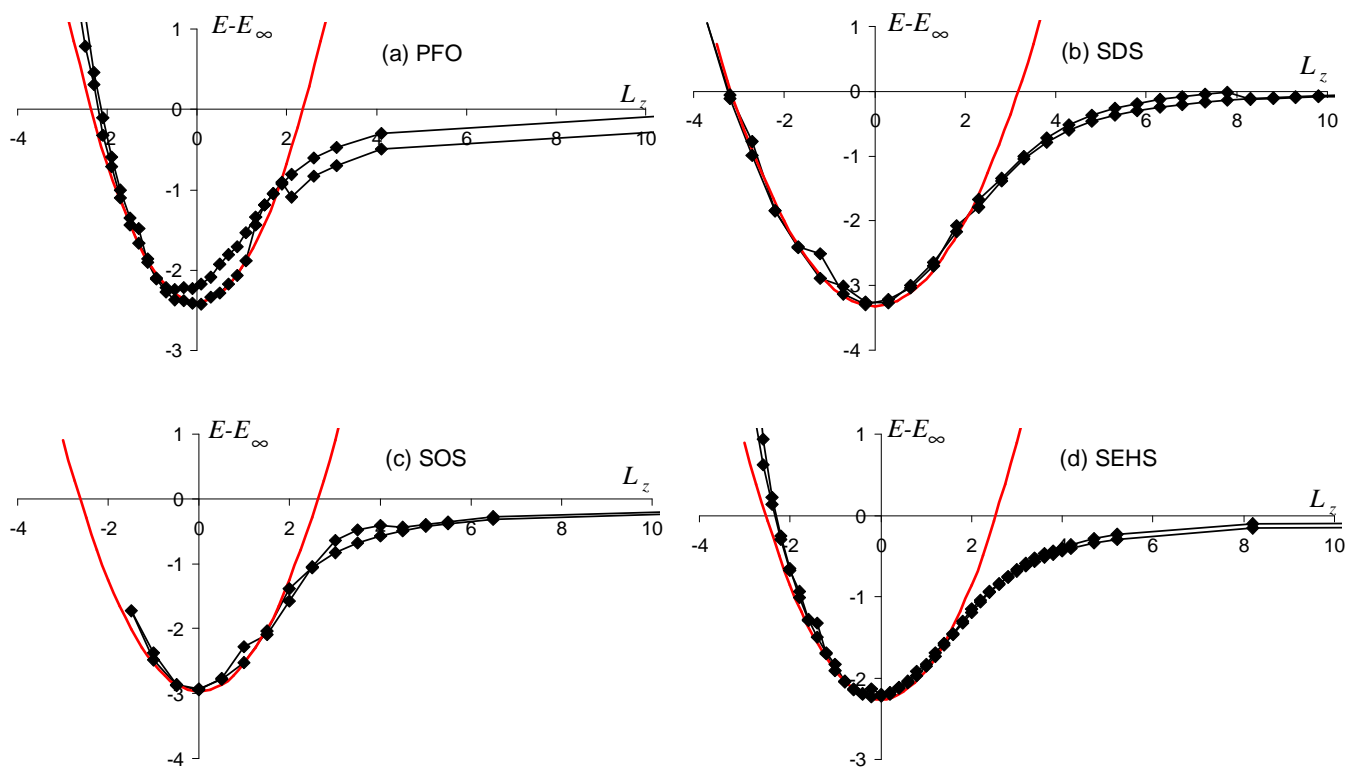

**Fig. 9.** Energy difference $E(L_z)$–$E(\infty)$ [kcal/mol] as a function of $L_z$ [Å] at the last cycle for (a) sodium perfluorooctanoate (PFO), (b) sodium dodecyl sulfate (SDS), (c) sodium 2-ethylhexyl sulfate (SEHS) and (d) sodium octyl sulfate (SOS). Red lines denote the fitting the energy difference by the parabolic potential.

For model validation, we calculated $\sigma_{cmc}$ for four simple anionic surfactants: sodium dodecyl sulfate (SDS), sodium octyl sulfate (SOS), sodium 2-ethylhexyl sulfate (SEHS) and sodium perfluorooctanoate (PFO) and compared the results to experimental data. The calculated energy dependence on $L_z$ for these surfactants is shown in Fig. 9 for the last cycles of ICO procedure along with the results of the fitting of $E(L_z)$ near the minimum by parabolic potential. The predicted $\sigma_{cmc}$ is compared to experimental data in Fig. 10 and Table 1. Table 1 also contains the calculated values of $a_{cmc}$, $\Delta\varepsilon$ and $\kappa$ for these surfactants. The obtained results are in a satisfactory agreement with experimental data (with accuracy better or equal 30%). We understand that proposed ICO procedure is imperfect and other atomistic approaches, e.g. Molecular Monte Carlo simulation, can be used for atomistic calculation of $\Delta\varepsilon$ and $\kappa$. However, we believe that obtained accuracy of $\sigma_{cmc}$ indicates that the model takes into account the major physics that governs $\sigma_{cmc}$, and ICO is applicable for the fast analytical screening of the possible candidates for the synthesis of new surfactant molecules.

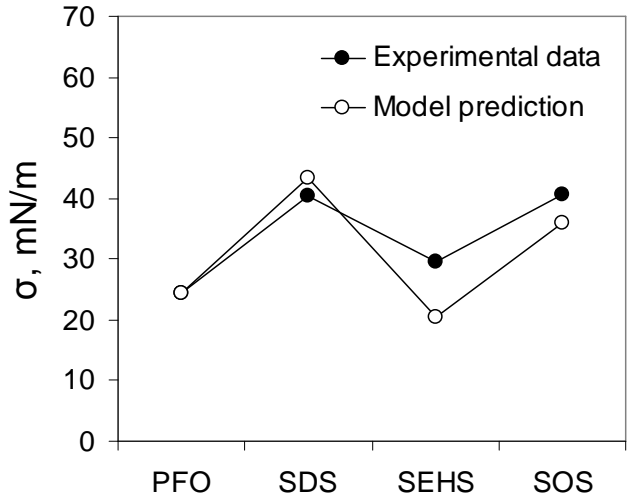

**Figure 10.** Calculated and experimental data for $\sigma_{cmc}$ for four selected surfactants.

**Table 1.** Experimental and calculated values of $a_{cmc}$, calculated energy difference $\Delta\varepsilon$, experimental and calculated surface tension at cmc point.

| Surfactant | Experimental | | Calculated | | | |
| --- | --- | --- | --- | --- | --- | --- |
| | $\sigma_{cmc}$, mN/m | Ref | $a_{cmc}$, Å$^2$ | $\Delta\varepsilon$, kcal/mol | Stiffness parameter $\kappa$, mN/m | $\sigma_{cmc}$, mN/m |
| PFO | 24.5 | [33] | 41.6 | 2.4 | 600 | 24.4 |
| SDS | 40.5 | [34] | 41.8 | 3.32 | 460 | 43.5 |
| SEHS | 29.5 | UTC internal exp. data | 45.2 | 2.26 | 490 | 20.5 |
| SOS | 40.7 | [34] | 41.5 | 2.97 | 600 | 35.9 |

## 5. Discussion and conclusions

A novel model for the surface tension calculation of surfactant solutions above the cmc point, $\sigma_{cmc}$, was developed. The model calculates $\sigma_{cmc}$ as a function of the chemical structure and composition of the surfactant molecule. The model takes advantage of equilibrium between the surfactant molecules at the liquid/vacuum interface (surface) and in the micelles in the bulk of the solution at cmc. The surface tension at equilibrium is equal to the difference in the free energy of the surfactant molecule at the surface and in the micelle, divided by the surface area per one surfactant at the surface.

Since the atomistic calculation of the free energy difference is a challenging problem, we derived an approximate analytical equation for $\sigma_{cmc}$, Eq. (26), based on the bilayer sheet approximation. The first, energetic, term in the right hand side of Eq. (26) is equal to the density of the energy of the cross-micelle interaction between two surfactant layers that form the bilayer. The second, entropic, term results from the restrictions on the surfactant molecule motion in the micelle imposed by the cross-micelle interaction. The Eq. (26) gives the lower estimate for $\sigma_{cmc}$ for given surfactant. As follows from Eq. (6) the surface tension decreases with increase of the free energy of surfactant in the micelle. The bilayer sheet approximation was utilized for derivation of Eq. (26). If any other types of micelle (spherical, cylindrical etc.) have lower free energy per surfactant, then these micelles are in equilibrium with the surfactants at the surface at lower surfactant concentrations than the bilayer formation. The formation of these micelles results in higher $\sigma_{cmc}$ than it would be if the bilayer formed at lower concentrations as follows from Eq. (6) (because these micelles have the lower free energy than the bilayer). Therefore, the actual surface tension would be higher than that calculated from Eq. (26) comparing the surface with the bilayer. The equation contains three parameters: the surface area per surfactant molecule, the density of the energy of the cross-micelle interaction of the layers in the bilayer, and the parameter characterizing the stiffness of the cross-micelle interaction. All these parameters were calculated utilizing a new atomistic approach.

The developed atomistic model exploits the similarity of the surfactant configuration on the surface and in the bilayer to reduce the impact of thermodynamic fluctuations. We start the simulation by the modeling of two isolated surfaces. Subsequently we gradually reduce the distance between the top and the bottom surfaces to obtain the bilayer model. That enables maintaining water subsystem in the same local minima in both isolated surfaces and bilayer systems in the calculation procedure while changing mostly the state of non-polar part of surfactant molecule. Moreover, $\Delta\varepsilon$ was calculated by Molecular Mechanics to avoid thermodynamic fluctuations of water molecules inherent in Molecular Dynamics. To provide the tail terminating groups with higher ability to reach the favorable configuration we cycled the system with respect to the distance between the top and the bottom layers of the bilayer. The calculation of $\sigma_{cmc}$ for several simple surfactants (SOS, SDS, SEHS, and PFO) showed reasonable accuracy (~30%). The agreement of the model prediction with experiment is better than one might expect from relatively small system size and CPU resource used in calculations.

The recommendations for design of surfactants with low $\sigma_{cmc}$ follow from the physical insight gained from the model. As follows from Eq. (6), $\sigma_{cmc}$ decreases with decrease of the difference between the free energy density of the surfactant at the surface and in the micelle, $\Delta\varepsilon$. The major contribution into $\Delta\varepsilon$ comes from the density of the energy of additional tail-tail interaction in the micelle. This interaction does not exist at the surface. Reduction of $\sigma_{cmc}$ can be archived by design of surfactant molecule with low energy density of this additional interaction in the micelle. That can be archived by the choice of the surfactant tails, which interact with each other only through short-range van der Waals interaction. Low tail-tail penetration in the micelle (see Fig. 3c) also decreases $\sigma_{cmc}$.

We assume that the tail-tail penetration is the smallest in the bilayer type of the micelle with compact placement of the tails (see Fig. 3b). Other types of the micelles (spherical, cylindrical etc.) result in higher tail-tail penetration than that in the bilayer. We speculate that sufficiently long and rigid tail suppresses formation of non-flat shape micelles and results in preferable formation of bilayer shape micelles. To prevent cross-micelle penetration of the tails the hydrophobic tail of the surfactant should be thicker than the hydrophilic head. In this case, the cross-micelle interaction is mostly governed by the van der Waals cross-micelle tail-tail interaction between the tail terminating groups belonging to the top and the bottom layers of the micelle.

The thermodynamic models of micellization in the surfactant solution were developed in Refs.[35, 36]. These models calculate the contributions of interactions in the micelle to the chemical potential of surfactant molecule. This enables to predict successfully the critical micelle concentration for different types of surfactants. However, the calculation of surface tension at the cmc requires the value of the chemical potential of surfactant molecule at the surface as and input in addition to the calculated chemical potential of surfactant molecule in the micelle. That requires the calibration of the model through experimental data [37]. The model presented in this paper does not have adjustable parameters and therefore does not require experiment calibration. That might be important for understanding and design of new surfactants for which experimental data are not available. Kinetics of surfactant absorption at the water/gas interface plays an important role in foam formation from aqueous solution of surfactants. The fundamental model for kinetics of surfactant adsorption and accompanying dynamics of surface tension was developed in Refs. [38, 39, 40]. The results presented in the current paper can be also utilized through providing the equilibrium surface tension to compliment the kinetic model developed in [38, 39, 40].

**Acknowledgements**

The authors gratefully acknowledge Dr. M. McQuade, Dr. D. Parekh, and Dr. M. Atalla of UTC and Professor J.M. Deutch of Massachusetts Institute of Technology and Professor G.M. Whitesides of Harvard University for the interest to the work, inspiring discussion and support.

**Figure and table captions**

**Fig. 1.** (a) Schematic representation of typical surface tension isotherm of water/surfactant solution. (b) Surface concentration of surfactant molecules, Γ. (c) Dependence of surface tension of water/surfactant solution on Γ.

**Fig. 2.** Schematic representation of reduction of the surface area and transfer of small Subsystem containing one surfactant molecule and the adjacent water molecules from the surface to the micelle Subsystem is highlighted in green.

**Fig. 3.** Schematic representation of: (a) two isolated surfaces (liquid/vacuum interfaces), and bilayer (b) without and (c) with cross-micelle tail-tail penetration.

**Fig. 4.** Configurations of surfactants (a) on liquid/vacuum surface and (b) in the bilayer.

**Fig. 5.** Schematic representation of weak perturbation in the form of triangle wave with small angle θ on the surface and the corresponding configuration of surfactant molecules.

**Fig. 6.** Numerical implementation of the model. We used periodic boundary conditions in $x$, $y$ and $z$ directions. The micelle model (a) was generated from the surface model (b) by decrease of $L_z$. Artificial non-physical interactions are indicated by the red arrowhead line.

**Fig. 7**. Cartoon of the potential energy landscape of two isolated surfaces (solid line) and bilayer (dashed line). Two minima (A and B) differentiate from each other by configurations of the water molecules. At the fixed arbitrary coordinate the isolated surfaces and bilayer have the same configurations of the water molecules and differentiate from each other only by the distance between the layers, $L_z$. The energy difference between the isolated surfaces and bilayer, Δε, which governs the $σ_{cmc}$, is much smaller than the energy difference between minima A and B, ΔE.

**Fig. 8.** (a) The energy of the system as a function of $L_z$ at the first and the last cycles and (b) the calculated energy of the "surface", "micelle" and the energy difference Δε for sodium perfluorooctanoate (PFO).

**Fig. 9.** Energy difference $E(L_z)$–$E(∞)$ [kcal/mol] as a function of $L_z$[Å] at the last cycle for (a) sodium perfluorooctanoate (PFO), (b) sodium dodecyl sulfate (SDS), (c) sodium 2-ethylhexyl sulfate (SEHS) and (d) sodium octyl sulfate (SOS). Red lines denote the fitting the energy difference by the parabolic potential.

**Figure 10.** Calculated and experimental data for $σ_{cmc}$ for four selected surfactants.

**Table 1.** Experimental and calculated values of $a_{cmc}$, calculated energy difference Δε, experimental and calculated surface tension at cmc point.